  \documentclass[final,5p,times,twocolumn]{elsarticle}

\usepackage{amssymb}

\usepackage{amssymb} 
\usepackage{algorithm} 
\usepackage{caption}
\usepackage{hyperref}
\usepackage{subcaption}
\usepackage{tabularx}
\usepackage{array}
\usepackage{color}
\usepackage{graphicx}
\usepackage{booktabs}
\usepackage{float}
\usepackage{url}
\usepackage{xcolor}
\usepackage{siunitx}  
\usepackage{multirow}
\journal{.}

\begin{document}

\begin{frontmatter}



\title{Abusive music transformation  using GenAI and LLMs}
 
\author[1,2]{Jiyong Choi}

\author[1,2]{Rohitash Chandra}

\affiliation[1]{Transitional Artificial Intelligence Research Group, School of Mathematics and Statistics, UNSW Sydney, Australia}
\affiliation[2]{Centre for Artificial Intelligence and Innovation, Pingla Institute, Sydney, Australia}

\begin{abstract}

Repeated exposure to violence and abusive content in music and song content can influence listeners' emotions and behaviours, potentially normalising aggression or reinforcing harmful stereotypes.
In this study, we explore the use of generative artificial intelligence (GenAI) to automatically transform abusive words (vocal delivery) and lyrical content in popular music. Rather than simply muting or replacing a single word, our approach transforms the tone, intensity, and sentiment, thus not altering just the lyrics, but how it is expressed. We present a comparative analysis of four selected English songs and their Gen-AI-transformed counterparts, evaluating changes through both acoustic and sentiment-based lenses. Our findings indicate that Gen-AI significantly reduces vocal aggressiveness, with acoustic analysis showing improvements in Harmonic to Noise Ratio, Cepstral Peak Prominence, and Shimmer. Sentiment analysis reduced aggression by 63.3-85.6\% across artists, with major improvements in chorus sections (up to 88.6\% reduction). The transformed versions maintained musical coherence while mitigating harmful content, offering a promising alternative to traditional content moderation that avoids triggering the "forbidden fruit" effect, where the censored content becomes more appealing simply because it is restricted. This approach demonstrates the potential for GenAI to create safer listening experiences while preserving artistic expression.

\end{abstract}


\end{frontmatter}

\section{Introduction}

 Music has become more accessible and widely consumed than ever before, with on-demand and Over-The-Top (OTT) streaming services \cite{ott}. At the same time, the ease of recording and uploading songs has become easier than ever, leading to an enormous amount of content available on the internet. Although this growth encourages growth and creativity, streaming platforms now face the challenges of detecting and mitigating harmful and inappropriate content while ensuring that freedom of expression is not limited.

Several studies have explored the influence and behaviours of individuals from songs that include lyrical content and music. Waite et al. \cite{waite1992mtv} found a significant decrease in the frequency (not severity) of aggressive behaviour on forensic patients after removing content from MTV (Music Television). Johnson et al. \cite{Johnson01021995} found that the participants showed greater acceptance of violence, a higher probability of engaging in violence, and greater acceptance of the use of violence against women for subjects who were exposed to violent rap videos compared to subjects who were not. A study in 2003 \cite{anderson2003violentlyrics} revealed that participants who heard a violent song felt more hostile and aggressive than those who heard a similar, but non-violent song. These studies suggest that the consumption of music with violent and abusive lyrics/videos is reflected in the listener's sentiment.

Natural language processing (NLP)\cite{otter2020survey, torfi2020natural}  is a branch of artificial intelligence for analysing and understanding natural language through text. It plays a vital role in applications such as sentiment analysis \cite{10.1371/journal.pone.0255615}, chatbots \cite{MISISCHIA2022421}, translation\cite{9715095} and abuse detection \cite{mishra2019tackling}.
Together with advancements in NLP, generative artificial intelligence(GenAI) \cite{gan, attention} has made impressive strides in music generation \cite{wavenet, musegan}. 
GenAI involves using deep learning models to generate new text, images, music, and more based on a given prompt. Notable examples include Large Language Models (LLMs) such as Generalised Pretrained Transformers (GPT) \cite{DBLP:journals/corr/abs-2005-14165} and Gemini \cite{geminiteam2025geminifamilyhighlycapable} for text, Midjourney \cite{midjourney} and Stable Diffusion \cite{rombach2022highresolutionimagesynthesislatent} for images, and tools such as Udio and Suno for music generation. Many of these tools, including GPT, DALL·E \cite{DBLP:journals/corr/abs-2102-12092}, and MusicGen \cite{copet2024simplecontrollablemusicgeneration}, are built on Transformer-based models.

Models such as  Meta's (Facebook) MusicGen, Suno, and Udio are reshaping how music is composed and consumed by generating high-quality, human-like music from minimal input \cite{music_gen_eval}. However, the rapid rise of these powerful generative systems has sparked significant ethical and legal controversies. A primary concern is the rise of DeepFakes \cite{deepfake}, i.e. AI-generated media that mimics real individuals, which have been used maliciously for scams, misinformation, and impersonation \cite{MUSTAK2023113368}. Furthermore, many models are trained on copyrighted material without appropriate consent, raising complex questions about legality and ownership. These controversies extend further, prompting ongoing debates on whether AI-generated art should be considered and appreciated as genuine artwork \cite{10.3389/fpsyg.2022.1024449}. Studies indicate a human bias against AI art, suggesting that human engagement in the artistic process contributes to appraisals of the art. \cite{aivshumanart}. These issues collectively underscore the urgent need for responsible AI development and rigorous research into the ethical, legal, and social implications of these increasingly capable and accessible systems. architectures.

Although streaming services such as Spotify have regulation services such as explicit tagging and human intervention, these tools have limited effects \cite{doi:10.1177/13548565251324508}, leaving many listeners with access to songs with extreme views and abusive comments. This failure is partly explained by psychological effects like the "forbidden fruit" theory, where warnings can make content more enticing by fuelling curiosity \cite{bushman}. An opposing "tainted fruit" theory suggests labels should steer consumers away, but studies on media ratings have found no consistent significant effects across different forms of media. For instance, Christensen's research on parental advisory labels found no significant increase in the attractiveness of music for adolescents, showing no support for the forbidden fruit effect \cite{parental_advisory}. This aligns with findings by Gosselt \cite{pictogram}, who discovered that age ratings did not make DVDs and video games more attractive to young audiences. In direct contrast, other studies have demonstrated that violent descriptions or warning labels can heighten interest in media like video games and television shows \cite{bushman, forbiddenfruitgaming}. This body of conflicting research reveals that the impact of advisory labels is not universal, varying based on the medium, audience, and type of content, which makes their effects difficult to predict.

There are many more issues with manual explicit tagging. For example, the label is voluntary \cite{RIAA2025ParentalAdvisory}, which can lead to confusion and there are no further alternatives for the user. The user must either listen to the harmful lyrics or avoid them. Additionally, the user misses out on instrumentals, melody, and other aspects of the music due to the harmful lyrics.


In this study, we focus on a novel application of GenAI for refining music and songs. We present a framework that replaces explicit vocals in music tracks with AI-generated, content-safe alternatives. Our approach aims to preserve the main elements of the song, i.e. melody and instrumentation, while altering only the problematic lyrics. Hence, our study makes three key contributions: (1) we frame the problem of harmful content mitigation as a generative transformation task rather than a detection and filtering problem, (2) we explore and leverage state-of-the-art GenAI models for vocal replacement, and (3) we evaluate each transformation using lyrical and audio analysis techniques through our framework. Our study investigates a potential pathway to enable safer listening experiences for adults and children, and applies machine learning for responsible content moderation on streaming platforms.

\section{Background}

\subsection{Music}
Before the digital era, music enthusiasts relied on physical formats such as Vinyl Records (LPs) \cite{vinyl}, which gave a major shift from broadcast radio in the early 1900s. LPs offered a more affordable and accessible way to enjoy music, which later transformed into cassette tapes and compact discs (CDs) with formats such as  MP3 players. Currently, streaming services such as Spotify and Apple Music make it easier and cheaper to listen to music on demand \cite{doi:10.1177/01634437231179347}. 

The concept of freely accessible music is not new, and became a significant debate with the rise of online sharing platforms. One of the earliest and influential examples of online music sharing was Napster \cite{napster} in the early 2000s. Napster was a Peer-to-Peer (P2P) file-sharing application that allowed users to share music files through the Internet. With over 60 million users sharing music, there was a significant shift in how people accessed and distributed music. As Napster grew, it attracted legal attention from artists such as heavy metal icons Metallica \cite{metallica-napster} and hip-hop artist Dr Dre. Eventually, in 2001, Napster was brought down by the Recording Industry Association of America (RIAA). Despite its downfall, the legacy of Napster played an essential role in shaping the future of digital music consumption, paving the way for the streaming services that we rely on today \cite{legacy}.

Although abusive lyrics in the Western world were quite common in the 80s with hair metal bands such as Motley Crue, and Guns and Roses \cite{gnr}, music from the 90s had extreme examples of lyrical abuse. Bands like Nirvana and Alice in Chains from the Seattle grunge movement, \cite{grunge}, and Hip Hop icons such as Eminem, Tupac Shakur (2Pac) and Snoop Dogg had many songs related to gang-affiliated abuse, threats, and more \cite{ffbfff32-f906-3a55-81c9-e2dc084de19a}.

\subsection{Language models and NLP}

One of the foundational machine learning models for handling sequential data is the Recurrent Neural Network (RNN) \cite{DBLP:journals/corr/abs-1808-03314} that feature feedback of information for processing long sequences of data, which  makes them well-suited for time-dependent data such as speech. However, RNNs struggled with long-term dependencies (\cite{REBER1967855}) due to the vanishing gradient problem. To address this, Long-Short Term Memory \cite{10.1162/neco.1997.9.8.1735} (LSTM) allows the neural network to forget and retain information, with gating mechanisms. LSTMs have proven to be effective even in recent years, \cite{sak14_interspeech}; however, they struggled with scalability in tasks with a very large amount of data. The development of the Transformer models \cite{DBLP:journals/corr/VaswaniSPUJGKP17}  eliminated the need for traditional recurrence by relying entirely on self-attention mechanisms in enhanced LSTM models. This allows them to process entire sequences in parallel and capture global dependencies more effectively for the development of a new wave of generative AI tools \cite{transformers_application}.

BERT (Bidirectional Encoder Representations from Transformers)\cite{devlin2019bertpretrainingdeepbidirectional} is a LLM (Large Language Model) developed by Google, which has made big strides in the field of NLP (Natural Language Processing). There are several prominent models that have been inspired by BERT, such as RoBERTa \cite{DBLP:journals/corr/abs-1907-11692}, ALBERT \cite{lan2020albertlitebertselfsupervised}, and DistilBERT \cite{sanh2020distilbertdistilledversionbert}. Unlike LLMs such as GPT \cite{openai2024gpt4technicalreport}, which are decoder-based models used specifically to generate new output sequences in response to an input sequence, BERT-based models are encoder-based only. The encoder block in the transformer architecture processes the input sequence and generates a rich numeric vector (or embedding) for each word. Encoder-only models stack multiple encoders to produce a single model. To fine-tune BERT for tasks such as question answering, sentiment analysis, and semantic analysis, linear layers are added on top of the stacked encoders, which are used to generate a specific input based on the task it is trying to solve. Another important feature of BERT models is it's bidirectional encoders, which allow for processing context on both sides of the word in the sentence at the same time. This is possible through MLM (Masked Language Modelling), where BERT randomly masks words and predicts them using all surrounding words. 

\section{Methodology}

\subsection{Data and Case Studies }

We focus our analysis on four selected  artists based on popularity and music content with abusive and explicit lyrics:

Kanye West (Ye) is one of the most influential hip-hop artists since the early 2000s, with record-breaking albums such as "The College Dropout", "Graduation" and "My Beautiful Dark Twisted Fantasy". Kanye has made a name for himself for his creative sampling, clean production and well-written lyrics. Kanye's lyrical input often featured a mix of cursing and sexual topics, but these were often considered harmless \cite{BurnsWoodsLafrance2016}, and many of his lyrics were perceived as optimistic. As an example, we refer to the first track of his third album, Graduation: "Good mornin', on this day we become legendary. Everything we dreamed of."
Although a prominent figure in the world of hip-hop, in February 2025, Ye publicly announced his anti-Semitic beliefs on Twitter (X) \cite{NYT2025YeTakesBackApology}. A few months later, he released a song named "Heil Hitler" (HH) on X, which many critics deem to be Kanye's worst song, as well as his most racist, and blocked from entering Australia \cite{BBC2025KanyeWestBlocked}.

Another prominent artist is rapper Cardi B, who gained popularity after the release of her debut album "Invasion of Privacy" in 2018. Her bold personality and evocative lyrics were highlighted with the release of "WAP" (Wet A** P****), which gained much attention from the media and internet due to its over-sexualisation and shallow lyrics. "WAP" sparked much debate about its lyrical content, and some listeners have regarded it as female empowerment, while many others find it to be vulgar and potentially harmful \cite{Williams2017CardiB}.

  Elton John \cite{Allcock2015EltonJohn} has been highly influential for decades with releases such as "Goodbye Yellow Brick Road", "Rocket Man", "Can You Feel the Love Tonight?" In recent years, his  "Cold Heart" was formed and remixed with pop singer Dua Lipa. In 1975, Elton John would release his single "Island Girl", a song about a Jamaican prostitute in New York City and a Jamaican man who wants to take her back to their home country. Despite its success, ranking number one for three weeks on the Billboard Hot 100 in the United States, the song gained much controversy due to its racially insensitive lyrics, and John has not performed it since 1990.

Tom MacDonald is a political rapper who rose to prominence in the late 2010s and early 2020s, largely through controversy and viral marketing. His music frequently targets subjects such as cancel culture, political correctness, and identity politics, but critics often accuse him of using oversimplified narratives to gain attention. Songs such as "Fake Woke,” “Snowflakes"” and “Brainwashed” have been criticised for promoting divisive rhetoric and capitalising on cultural tensions. Although he claims to speak uncomfortable truths, many listeners view his work as inflammatory, self-serving, and lacking genuine artistic depth. 

\subsection{Audio separation method}

HTDemucs (Hybrid Transformer Deep Extractor For Music Sources) \cite{rouard2023hybrid} is a deep learning model used for music source separation. The deep learning model has been trained on a popular dataset for music source separation, which included 150 tracks (100 for training and 50 for testing), each with a sample rate of 44.1 kilohertz. 
U-Net is a type of convolutional neural network originally designed for image segmentation \cite{ronneberger2015unetconvolutionalnetworksbiomedical} that consists of three main parts. The encoder converts input into a dense representation called an embedding, and the decoder upsamples the compressed representation back to the original input size. Skip connections directly connect earlier layers to later layers, bypassing the layers in between which allows the decoder to have access to both high-level and low-level information.

The original HT-Demucs model \cite{défossez2022hybridspectrogramwaveformsource} combines a time domain U-Net (1d Convolutions on waveform) and a spectrogram domain U-Net (2D convolutions on frequency time STFT). Each U-Net has 5 encoder layers and 5 decoder layers. After the 5th decoder layer, both branches produce feature maps of the same shape, which are summed and passed onto the 6th layer. The first decoder layer is shared, and the model splits back into the time and spectral decoders. The spectral output is converted back into a waveform using ISTFT (Inverse Short Term Fourier Transform) to create the final audio.
HT-Demucs \cite{rouard2022hybridtransformersmusicsource} keeps the outer 4 U-Net layers the same as the original hybrid architecture, but replaces the two innermost encoder and decoder layers with a cross-domain Transformer Encoder. 
The transformer processes the waveform and spectrogram in parallel and alternates self-attention and cross-attention layers between the two domains.



\subsection{Music generation using Suno AI}

Suno AI  \cite{Suno2025}  is a music generation platform that converts text into audio (songs).
It features a Transformer-based encoder that processes the user’s prompt, such as lyrics and musical genre, and converts it into a conditioning embedding that represents the mood, genre, and structure. This embedding is then passed to a large Transformer, which generates a blueprint of chord progressions, melody, and structure. The blueprint and text conditioning are used to generate a dense latent audio representation (i.e. spectrogram) through a latent diffusion model (LDM). The diffusion model \cite{ho2020denoisingdiffusionprobabilisticmodels} iteratively denoises the latent audio to produce clear vocals and instrumentation, and finally, a neural vocoder converts the spectrogram into a waveform. We use Suno AI as our main tool for transforming the vocals.

Suno AI allows users to upload parts of audio up to one minute, where the user can cover, extend or reuse the audio. The cover alters the song's style while maintaining a similar melody and form. Extend retains the original sound and creates something new from a selected point, while Reuse generates a new song using the same prompt and style as a previous one. Our main focus is to use the "cover" function to generate vocals over the instrumental track we provide. Suno AI also allows the user to input lyrics and styles. To rewrite the lyrics in a more positive tone, we employ GPT-4 with the following prompt: \textit{"Rewrite the lyrics so that it is not abusive and make sure it has the same length and flow: [lyrics]".} We use the output and feed the result into Suno AI. We use the style input to enter tags for the type of style; e.g., "trap", "hip-hop" will generate a song in such a style. We make use of this to generate a track that is as close to the original track as possible, with only the change in vocals to avoid genre changes and different vocal styles.

\subsection{Transformation and Evaluation Frameworks}

We present two frameworks that feature music transformation and transformed music evaluation.

\subsubsection{Music Transformation Framework}

\begin{figure*}[htbp]
    \centering
    \includegraphics[width=1\linewidth]{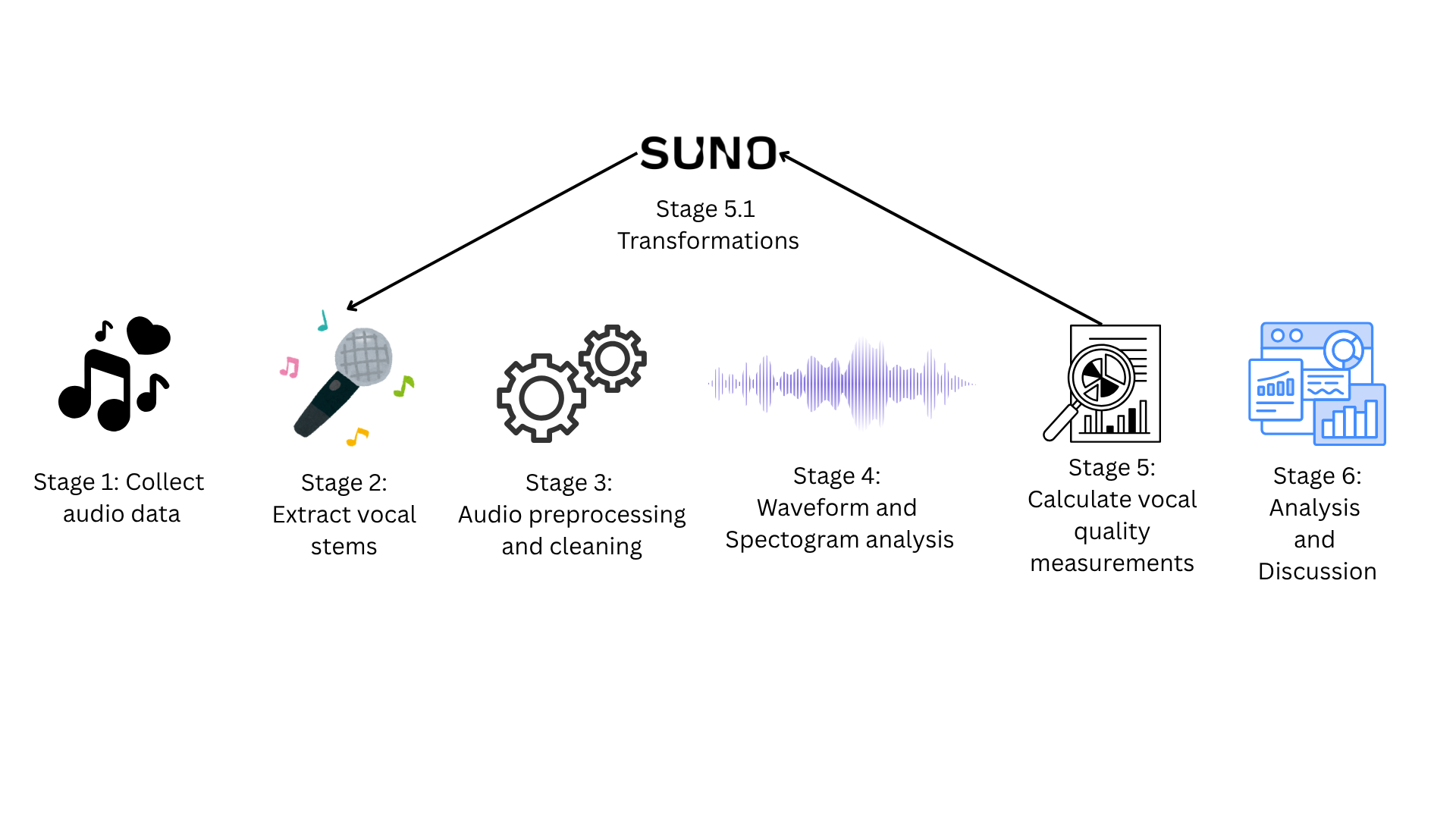}
    \caption{Music Transformation Framework showing stages including audio data collection and pre-processing, waveform and spectrogram analysis. We then implement vocal quality measurements and use the Suno AI to replace vocals, transforming the song. }
    \label{fig:rms}
\end{figure*}

\begin{figure*}[htbp]
    \centering
    \includegraphics[width=1\linewidth]{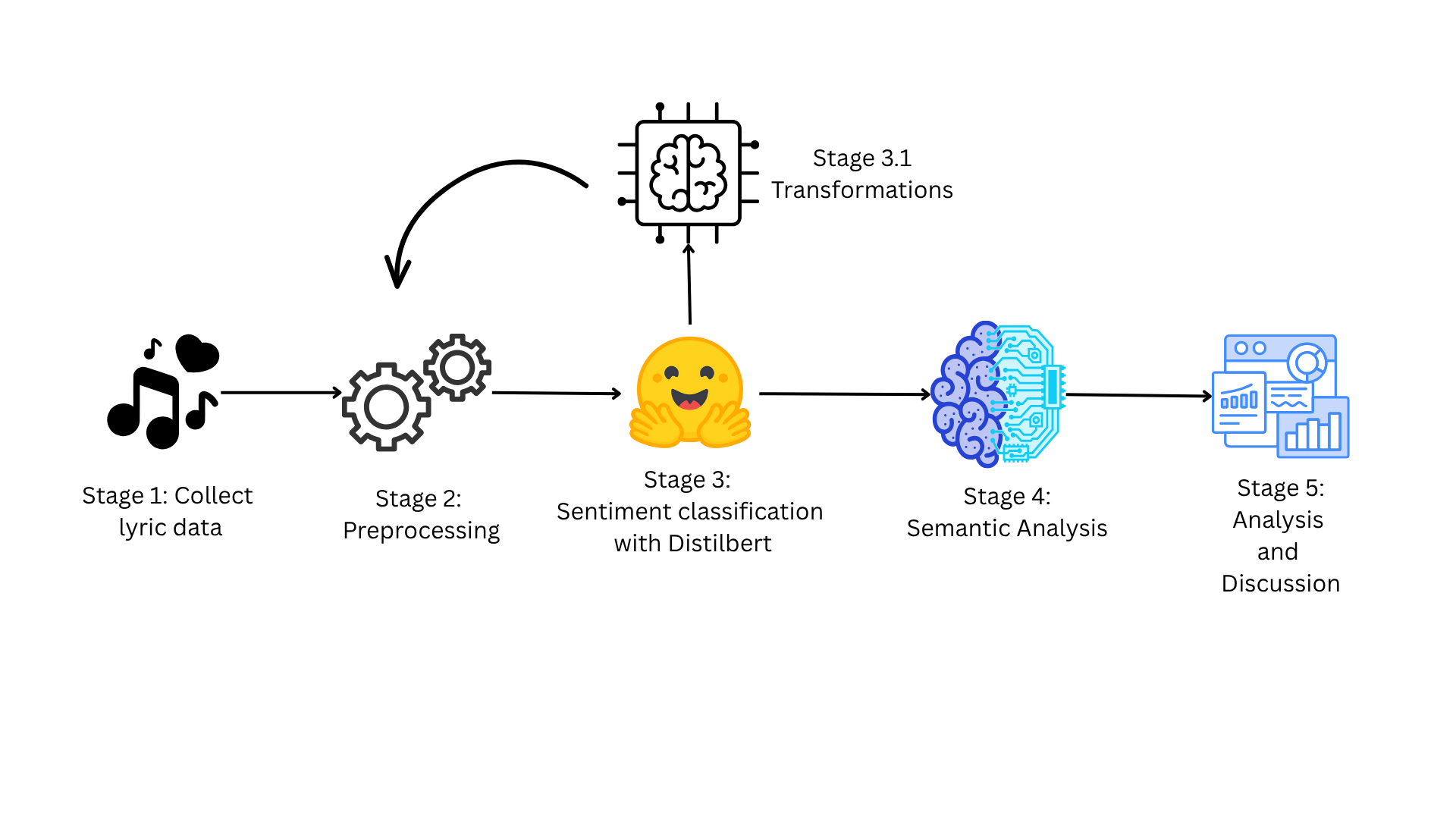}
    \caption{Music Evaluation Framework showing major stages that include collecting lyrical data, pre-processing data and sentiment analysis. The sentiment analysis pre and post  transformation of the lyrics evaluates if sentiments are maintained in transformed songs. We also calculate the cosine similarity between the original and transformed songs for semantic analysis. }
    \label{fig:rms}
\end{figure*}


The framework consists of two main stages, including the transformation of audio and lyrics using deep learning models. In Stage 1 of the audio framework, we download MP3 files for each of the four chosen songs. In Stage 2, we use audio separation techniques (Demucs \footnote{\url{https://github.com/facebookresearch/demucs}}) to separate the vocals from the instrumentals/beat.
Demucs is distinguished by its hybrid encoder-decoder architecture. The encoder uses convolutional layers to transform the raw audio waveform into a latent representation, capturing hierarchical features. A bidirectional LSTM \cite{6795963} (Long Short-Term Memory) layer at the bottleneck of this architecture models sequential dependencies across the entire track. Finally, a convolutional decoder reconstructs the separated source waveforms from this processed representation. For our purposes, we specifically extract the vocals stem for lyrical analysis \cite{DBLP:journals/corr/OSheaN15}.

Stage 3 involves preprocessing the audio, where the preprocessing steps include maintaining a uniform sampling rate, applying a pre-emphasis transformation to make up for spectral tilt, and remove low frequency noise. 
We preprocess the audio, then analyse the vocals. We utilise Suno AI to generate songs similar to the original with replaced vocals and lyrics, and perform the same analysis (lyrical and vocal). We make use of audio analysis metrics such as Harmonics to Noise ratio \cite{hnr}, and Cepstral Peak Prominence \cite{cpp}to evaluate the emotions of the vocals, and discuss our findings and insights.

In Stage 4, we plot the preprocessed waveform and spectrograms. The waveform plot shows the energy (db) of the audio signal over time. It helps us identify which parts of the songs are loud (a common trait for aggressive songs) and gives us a clear idea of how the song is structured. The spectrogram analysis shows the frequency (Hz) of the audio signal over time and indicates loudness depending on vocals and musical instruments. For example, drums and bass show in low frequencies and vocals and cymbals show in higher frequencies. In Stage 5, we evaluate the vocal quality using specialised metrics.

\subsubsection{Music Evaluation Framework}

For Stage 1 of the lyrical framework, we collect lyrics from websites such as Genius and LyricFind and save them as a data file. In Stage 2, we preprocess the lyrics for each artist, which includes removing section labels such as "[Chorus]" and "[Verse]", lowercasing, removing special characters, and stopwords (common words like "the" and "a").

In Stage 3, we perform sentiment analysis classification with DistilBERT via the Transformer library \footnote{  DistilBERT-base-uncased-finetuned-sst-2-english model\url{}}.  The DistilBERT \cite{sanh2020distilbertdistilledversionbert} model is a distilled \cite{hsieh2023distillingstepbystepoutperforminglarger} version of the BERT  model that performs almost up to the original model, but is significantly faster rate while taking lower memory. This is achieved through a knowledge distillation process during pre-training, where a smaller student model (DistilBERT) is trained to replicate the behaviour of a larger teacher model (BERT). The model has been further fine-tuned on the Stanford Sentiment Treebank v2 (SST-2) dataset \cite{socher-etal-2013-recursive}. The sentiment function returns a dictionary with key: label and value: "POSITIVE" or "NEGATIVE", and a key: score and value a number ranging from 0 to 1, representing the model's confidence in its prediction.

The label indicates a positive or negative sentiment, whereas the score value represents the strength and intensity of the provided text.
For example, "f*ck you" will output the following: \texttt{[{'label': 'NEGATIVE', 'score': 0.992}]}, whereas "thank you" will output \texttt{[{'label': 'POSITIVE', 'score': 0.999}]}. As a high score can either mean a strong positive or a strongly negative sentiment, we standardise the values and define the sentiment score with the following formula:
\texttt{standardised = 0.5 - (score-0.5) if label is "POSITIVE" else standardised = 0.5 + (score-0.5)}.

In Stage 3, we use GPT-4 to transform the lyrical content into a positive tone. We use the following prompt: \textit{"Rewrite the lyrics so that it is not abusive and make sure it has the same length and flow: [lyrics]".} We then repeat stages 2 and 3 for the transformed lyrics. We implement Semantic Analysis in Stage 4, where we use the MPNET model \cite{} to extract embeddings from each lyric. We perform cosine similarity between the original and transformed versions of the lyrics.
Finally, in Stage 5, we provide analysis and discuss whether lyrical transformation can help reduce the explicit and harmful nature of the songs.



\subsubsection{Music Analysis Metrics}

Yumoto \cite{yumoto1982} introduced  Harmonics to Noise Ratio (HNR), which has been widely established as a measure of vocal roughness and hoarseness \cite{anikin2017}, with lower values indicating roughness.
 
\begin{equation}
\text{HNR} = 10 \cdot \log_{10} \left( \frac{\sum_{h=1}^{H} E_h}{E_{\text{noise}}} \right)
\end{equation}
where $E_h$ is the energy at harmonic frequency $h \cdot f_0$, $H$ is the number of harmonics, and $E_{\text{noise}}$ is the residual noise energy.

The Cepstral Peak Prominence (CPP) measures the clarity and regularity of vocal fold vibration by quantifying the dominant cepstral peak relative to the cepstral baseline. A high CPP indicates clear periodic measure; e.g., a metronome has a high CPP, whereas popcorn popping is irregular with random timings, so it has a low CPP. The same can be applied to vocals; a high CPP indicates clear singing vocals, whereas a low CPP indicates hoarseness and vocal strain.

\begin{equation}
\text{CPP} = \max(\mathbf{c}[q]) - \frac{1}{Q} \sum_{q=1}^{Q} \mathbf{c}[q]
\end{equation}
where $\mathbf{c}[q]$ is the cepstrum at frequency $q$, and $Q$ is the number of frequency bins. The cepstrum is computed as:
\begin{equation}
\mathbf{c}[q] = \mathcal{F}^{-1} \left\{ \log \left( |\mathcal{F}\{x[n]\}|^2 \right) \right\}
\end{equation}
where $\mathcal{F}$ denotes the Fourier transform and $x[n]$ is the audio frame.

Jitter is the cycle-to-cycle variability in the fundamental frequency F0 of the voice. Higher jitter values indicate greater pitch instability, often associated with physiological tension or heightened emotional response. 
\begin{equation}
\text{Jitter (\%)} = \frac{1}{N-1} \sum_{i=1}^{N-1} \frac{|T_i - T_{i+1}|}{\overline{T}} \times 100
\end{equation}

where \(T_i\) is the duration of the \(i\)-th pitch period, \(\overline{T}\) is the mean pitch period, and \(N\) is the total number of pitch periods.

Shimmer refers to cycle-to-cycle variability in the amplitude (loudness) of the voice. Elevated shimmer reflects instability in vocal intensity and can be related to emotional expressions and stress in the vocal folds. Shimmer quantifies cycle-to-cycle variability in the amplitude of the voice signal, calculated as:

\begin{equation}
\text{Shimmer (\%)} = \frac{1}{N-1} \sum_{i=1}^{N-1} \frac{|A_i - A_{i+1}|}{\overline{A}} \times 100
\end{equation}

where \(A_i\) is the peak-to-peak amplitude of the \(i\)-th pitch period, \(\overline{A}\) is the mean amplitude, and \(N\) is the total number of periods.

\section{Results}
\subsection{Setup}

We first gather four songs with aggressive, inappropriate, racist, or harmful tones and download their respective MP3 files along with their lyrics. We collect the following songs: "Heil Hitler" by Kanye West (Ye), "WAP" by Cardi B, "Island Girl" by Elton John and "Whiteboy" by Tom Macdonald.

We begin to read into the lyric text files and begin preprocessing. We remove stop words such as "the" and "a" as they contribute little to no semantic meaning. 

\subsection{Lyrics Analysis}

\begin{figure}[htbp]
    \centering
    \includegraphics[width=1\linewidth]{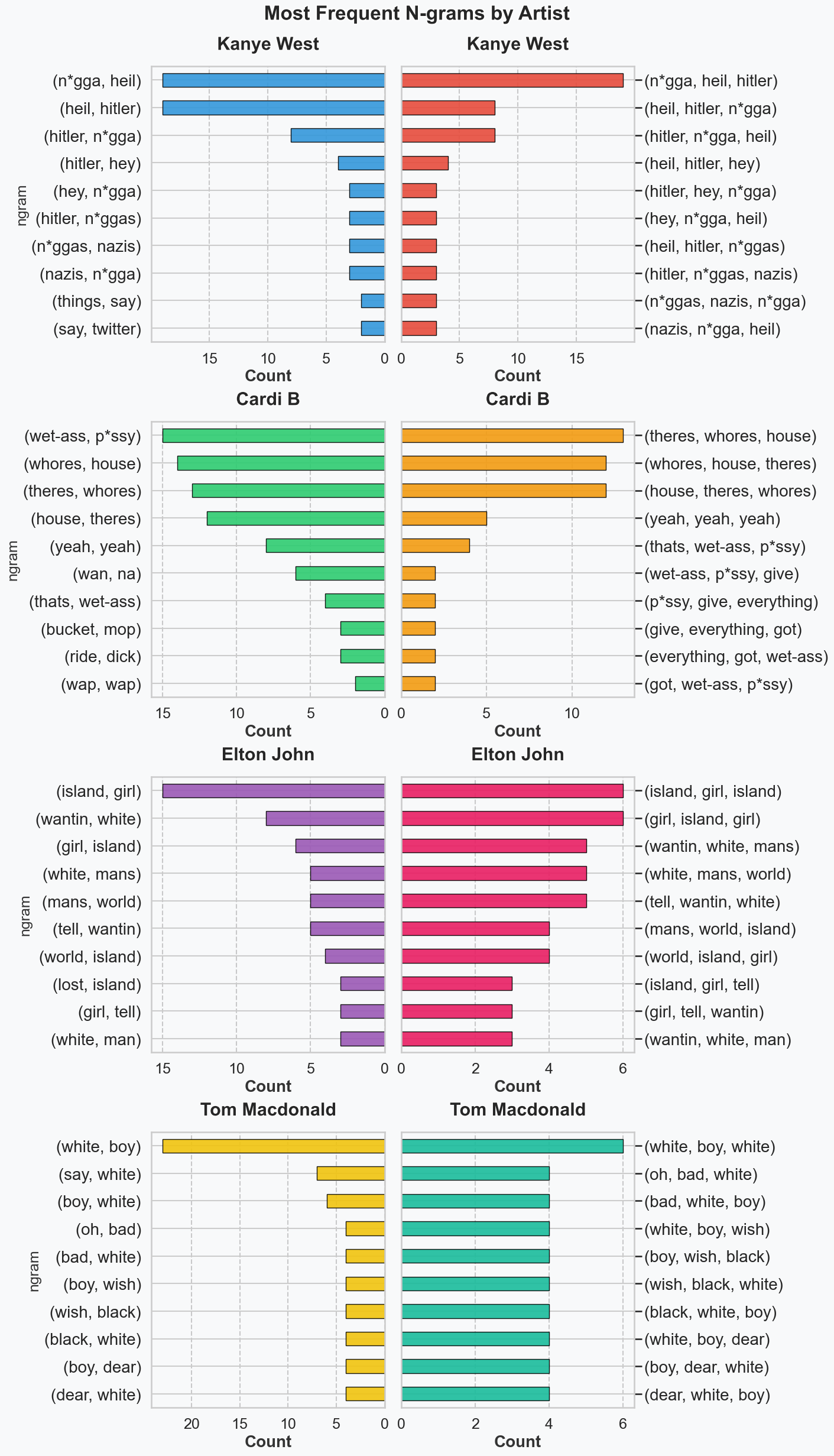}
    \caption{N-gram analysis of song lyrics showing the most frequent word sequences for each song.}
    \label{fig:ngram}
\end{figure}

We perform lyrical analysis for each of the songs to gain insight into the lyrical content of the abusive song. As shown in Figure \ref{fig:ngram}, the most frequent n-grams for each artist largely reflected repeated chorus lines. For example, Kanye West’s top bigram and trigram were “n*gga heil” and “n*gga heil hitler”, while Cardi B’s most frequent n-grams included “wet-*ss p*ssy” and “there’s wh*res house.” Examining these results, we find that Kanye West and Cardi B’s lyrics are explicitly abusive, where the words themselves are offensive. By contrast, Elton John and Tom Macdonald’s lyrics tend to be more implicitly abusive, requiring consideration of the broader song context to identify abusive meaning.

Table \ref{tab:sections} presents the average sentiment score by artist and song section. For Cardi B, we find negative sentiment most concentrated in the outro (0.998) and least in the verses (0.605). Elton John exhibits the lowest prevalence of negativity overall, with a mean sentiment score of 0.642 in the verse and only 0.157 in the chorus. Kanye West has the strongest negative sentiment on average for verse and chorus, with both skewing heavily negative (0.863 and 0.976). Tom Macdonald’s sentiment varied more widely, ranging from approximately 0.385 to 0.854 negative depending on the section.

\begin{figure*}[htbp]
    \centering
    \includegraphics[width=1\linewidth]{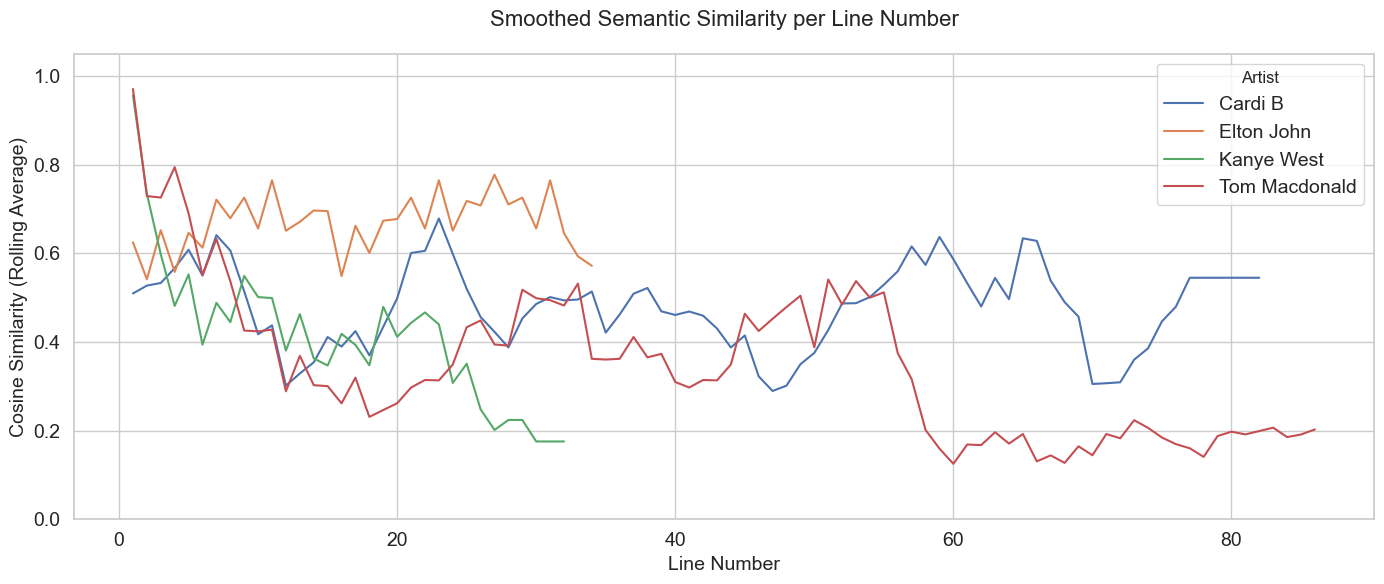}
    \caption{Line graph of cosine similarity between the original and transformed versions of text with a  rolling window size of 5.}
    \label{fig:semantic}
\end{figure*}

We extract embeddings for each line using mpnet-base-v2 and compute the cosine similarity between the original and transformed versions.
\ref{fig:semantic} shows the rolling average (window=5) cosine similarity for each artist. Elton John has an average similarity score, and the other artists fall just below 50\%. Kanye West and Tom Macdonald's songs reached down to 20\%, showing the transformations may not have preserved semantic meaning as well for these artists.

\begin{table}[htbp]
\centering
\caption{Average Semantic Similarity Between Original and Transformed Lyrics by artist}
\label{tab:semantic_similarity}
\begin{tabular}{l S[table-format=1.6]}
\toprule
\textbf{Artist} & \textbf{Semantic Similarity} \\
\midrule
Elton John     & 0.660\\
Cardi B        & 0.480\\
Kanye West     & 0.375\\
Tom Macdonald  & 0.333\\
\bottomrule
\end{tabular}
\end{table}

\subsection{Audio Analysis}

Although lyrics analysis gives us the content of the song, audio analysis reveals the context and emotion behind each line delivery. 
We employ  Demucs software to separate stems (vocals and instrumentals) from each song's MP3 file \cite{rouard2022hybridtransformersmusicsource}. Before this analysis, we preprocessed all vocal tracks to ensure feature consistency and reliability. We resampled each audio file to a uniform sampling rate of 22,050 Hz and truncated it to a maximum duration of 360 seconds. First, we employed pre-emphasis filtering to enhance high-frequency components and to flatten the spectrum. Since human vocals have a spectral tilt, meaning energy tends to decrease as frequency increases, we emphasise the high-frequency components with the following equation:  \[
H(z) = 1 - 0.97z^{-1}
\] as mentioned in \cite{liang2022extraction}. 

We implemented background noise reduction using the first 0.5 seconds of each track as a noise profile for spectral subtraction.
Furthermore, we added additional specialised processing to address artifacts in Demucs extracted vocals, including filtering to remove sounds under 100Hz and removing background noise by finding the quietest moments in the song, then subtracting them from the overall song. Finally, we implement amplitude normalisation to the range [-1, 1] to ensure that all tracks have comparable relative loudness for subsequent analysis.

\begin{figure*}[htbp]
    \centering
    \includegraphics[width=1\linewidth]{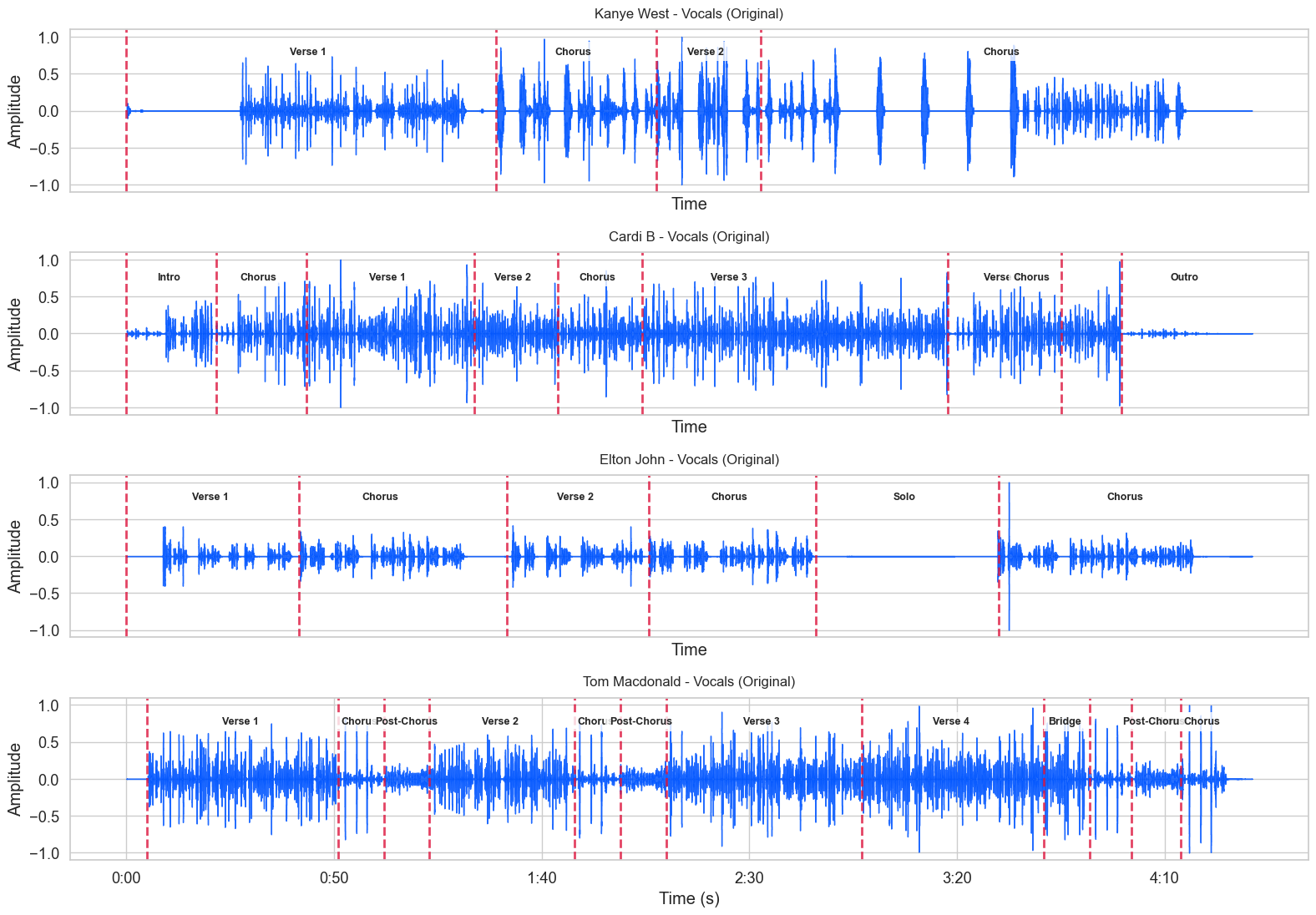}
    \caption{Waveform analysis separated by section}
    \label{fig:waveform}
\end{figure*}

\begin{figure*}[htbp]
    \centering
    \includegraphics[width=1\linewidth]{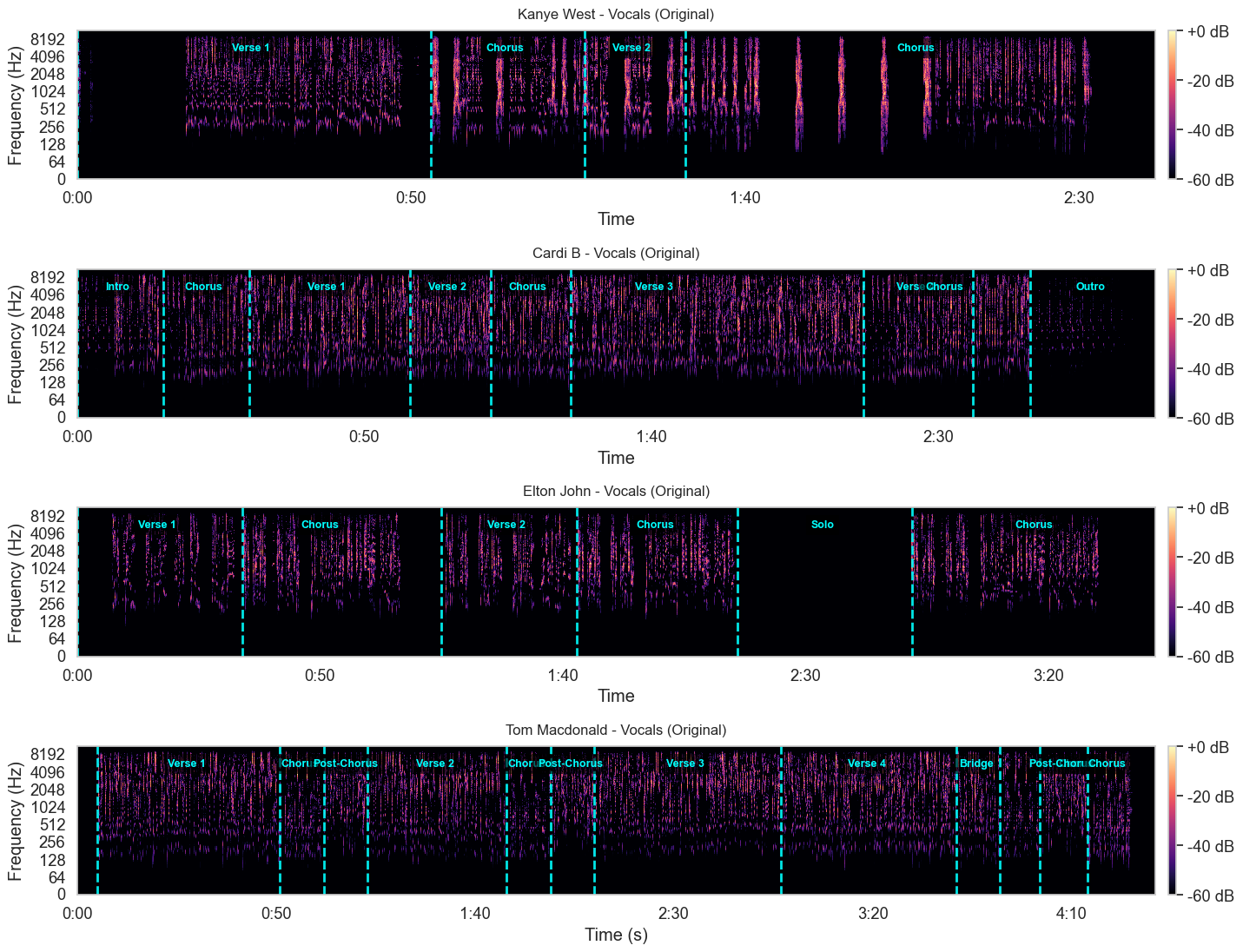}
    \caption{Spectrogram analysis separated by section}
    \label{fig:spectrogram}
\end{figure*}

We analyse each song using audio analysis techniques. We first plot   the waveform (Figure \ref{fig:waveform}) and spectrogram (Figure \ref{fig:spectrogram})  for each  song. The waveform shows the normalised amplitude (loudness), and the spectrogram, in contrast, shows the time-frequency representation. The spectrogram is generated by performing a series of STFT (Short Time Fourier Transform), which involves segmenting parts of the audio into windows, applying a Fourier transform and plotting the resulting spectrum of each window sequentially. We observe that Kanye’s vocal waveforms show distinct groupings with high amplitude peaks aligning with the chorus and moderate intensity through the verses. The spectrogram displays concentrated energy in the low–mid to high range (200-8000 Hz) and shows bright bands from 30 seconds onwards. The persistent high-frequency harmonics reinforce vocal tension and aggression. 

Furthermore, Cardi B’s waveform  (Figure \ref{fig:waveform})  shows consistent amplitude with few silent segments, revealing a steady, dominant vocal presence. The corresponding spectrogram highlights sustained mid- to high-frequency energy, showing bright areas in the spectrogram especially near verse 3. This steady energy pattern reflects verbal intensity associated with aggressive or commanding vocal delivery. Elton John’s vocal waveform  (Figure \ref{fig:waveform})  shows the quietest/low-energy waveforms compared to all other artists.  Spectrally, his vocals demonstrate strong, evenly spaced formants and a balanced distribution of energy across frequencies. The reduced high-frequency emphasis produces a warmer quality. The Suno AI transformation generally preserves these patterns but introduces a mild spectral sharpening, which slightly increases clarity at the expense of warmth, subtly shifting the tone toward modern brightness. Tom Macdonald’s waveform  (Figure \ref{fig:waveform})  features large dynamic fluctuations and strong peaks during choruses, reflecting heightened emotional intensity. The spectrogram shows brighter areas near the 4000-8000 Hz range, which reveals breathiness, tension and edge.


\subsection{Transformations}

\begin{figure}[htbp]
    \centering
    \includegraphics[width=1\linewidth]{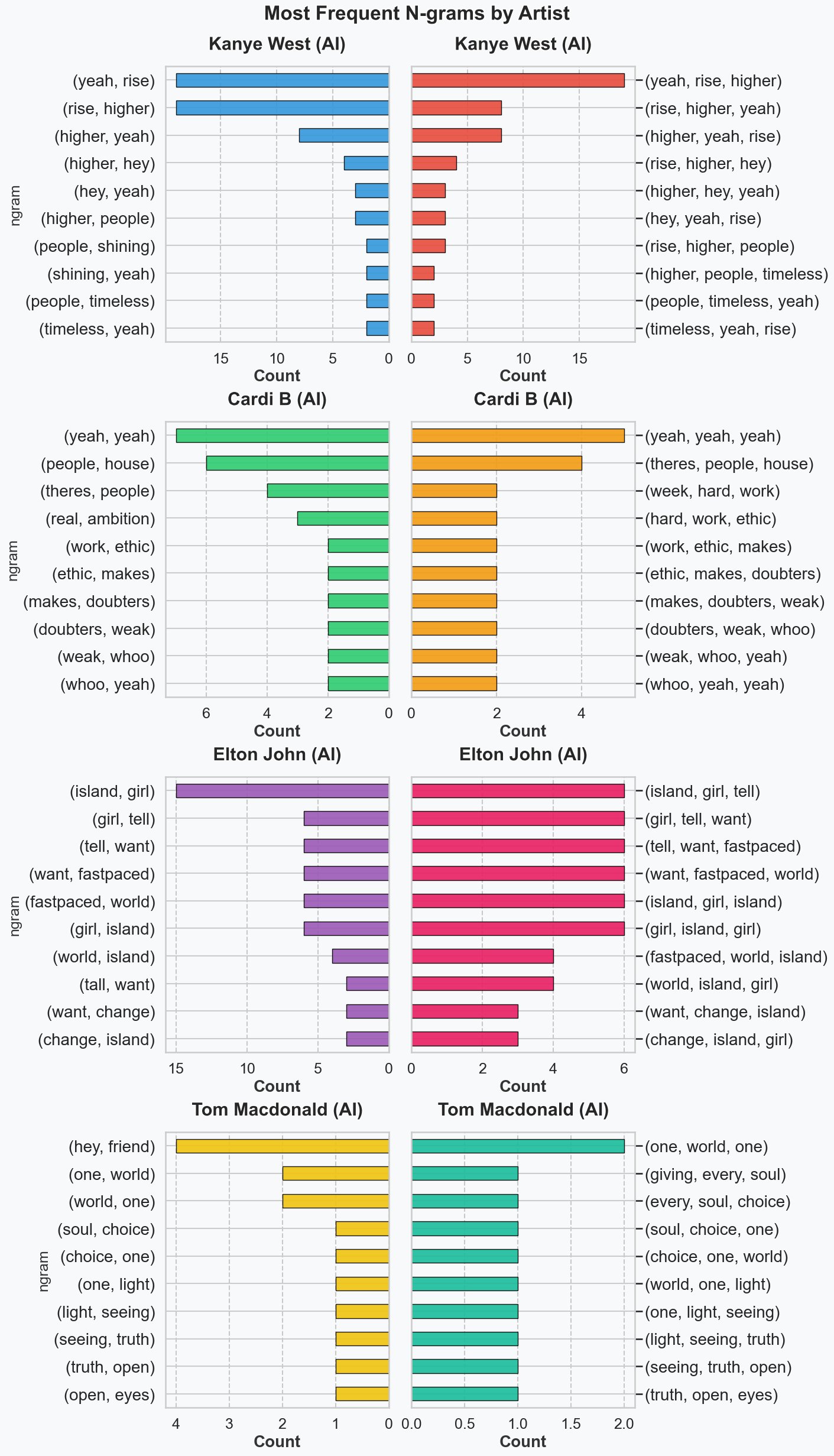}
    \caption{N-gram analysis for each song after transformation.}
    \label{fig:ngramAI}
\end{figure}

We analyse the bigram and trigram contents for each song to review the impact of the transformation. Our results in Figure  \ref{fig:ngramAI} show that after the transformations, each artist's most common n-grams included (rise, higher) instead of (heil, hitler) for Kanye. We observe (yeah, yeah, yeah) instead of (there's, whores, house) for Cardi B. Furthermore, we observe (hey, friend) instead of (white, boy) for Tom Macdonald's song. Finally, there was not much difference after the transformation for Elton John's song, as the lyrics were quite implicit and not immediately offensive.


We now perform sentiment analysis for each \textcolor{red}{song by defined sections.} In Table  \ref{tab:sectionsAI}, we observe that all transformed average sentiment scores decreased compared to the original results, indicating a positive transformation. In particular, Cardi B's inappropriate intro, chorus and outro sections reach almost 0\% in average sentiment score, similar to Elton John's song. Kanye West's HH did not see a major decrease in sentiment score for the verse, but did see a significant change in the chorus (from 0.976  to 0.111). 
We also examine the average sentiment by artist per line in the song lyrics. Table \ref{tab:sentiment_mean} shows a decrease in average sentiment for each artist, indicating that the transformation was successful. The percentage decrease was highest with Cardi b's song "WAP" at 85.6\%, and lowest with Kanye West's song at 63.3\%. Table \ref{tab:chorus} shows a side-by-side comparison between the original chorus and the transformed chorus.

\begin{table}[htbp]
\centering
\caption{Mean sentiment score by artist per line}
\label{tab:sentiment_mean}
\begin{tabular}{l S[table-format=1.6] S[table-format=1.6] S[table-format=1.6] S[table-format=2.6]}
\toprule
\textbf{Artist} & \textbf{Avg Sentiment} & \textbf{Avg Sentiment (AI)} & \textbf{\% Decrease}\\
\midrule
KW & 0.938 & 0.344 & 63.3 \\
CB & 0.744 & 0.107 & 85.6\\
EJ & 0.235 & 0.063 & 73.2\\
TM & 0.685 & 0.250 & 63.5\\
\bottomrule
\end{tabular}

\smallskip

\end{table}

\begin{table}[htbp]
\centering
\caption{Sentiment scores by section of the song. We use dashes to show that the particular song did not have a specific section.}
\label{tab:sections}
\begin{tabular}{lcccc}
\toprule
\textbf{Section} & \textbf{CB} & \textbf{EJ} & \textbf{KW} & \textbf{TM} \\
\midrule
Intro   & 0.884 & -     & -     & -     \\
Verse   & 0.605 & 0.642 & 0.863 & 0.611 \\
Chorus  & 0.794 & 0.157 & 0.976 & 0.854 \\
Bridge  & -     & -     & -     & 0.385 \\
Outro   & 0.998 & -     & -     & -     \\
\bottomrule
\end{tabular}
\end{table}

\begin{table}[htbp]
\centering
\caption{Sentiment scores by section of the song. We use dashes to show that the particular song did not have a specific section.}
\label{tab:sectionsAI}
\begin{tabular}{lcccc}
\toprule
\textbf{Section} & \textbf{CB (AI)} & \textbf{EJ (AI)} & \textbf{KW (AI)} & \textbf{TM (AI)} \\
\midrule
Intro   & 0.001002 & -        & -        & -        \\
Verse   & 0.265934 & 0.000521 & 0.608447 & 0.395731 \\
Chorus  & 0.005783 & 0.017889 & 0.111285 & 0.101181 \\
Bridge  & -        & -        & -        & 0.150715 \\
Outro   & 0.000340 & -        & -        & -        \\
\bottomrule
\end{tabular}
\end{table}

\begin{figure*}[htbp]
    \centering
    \includegraphics[width=1\linewidth]{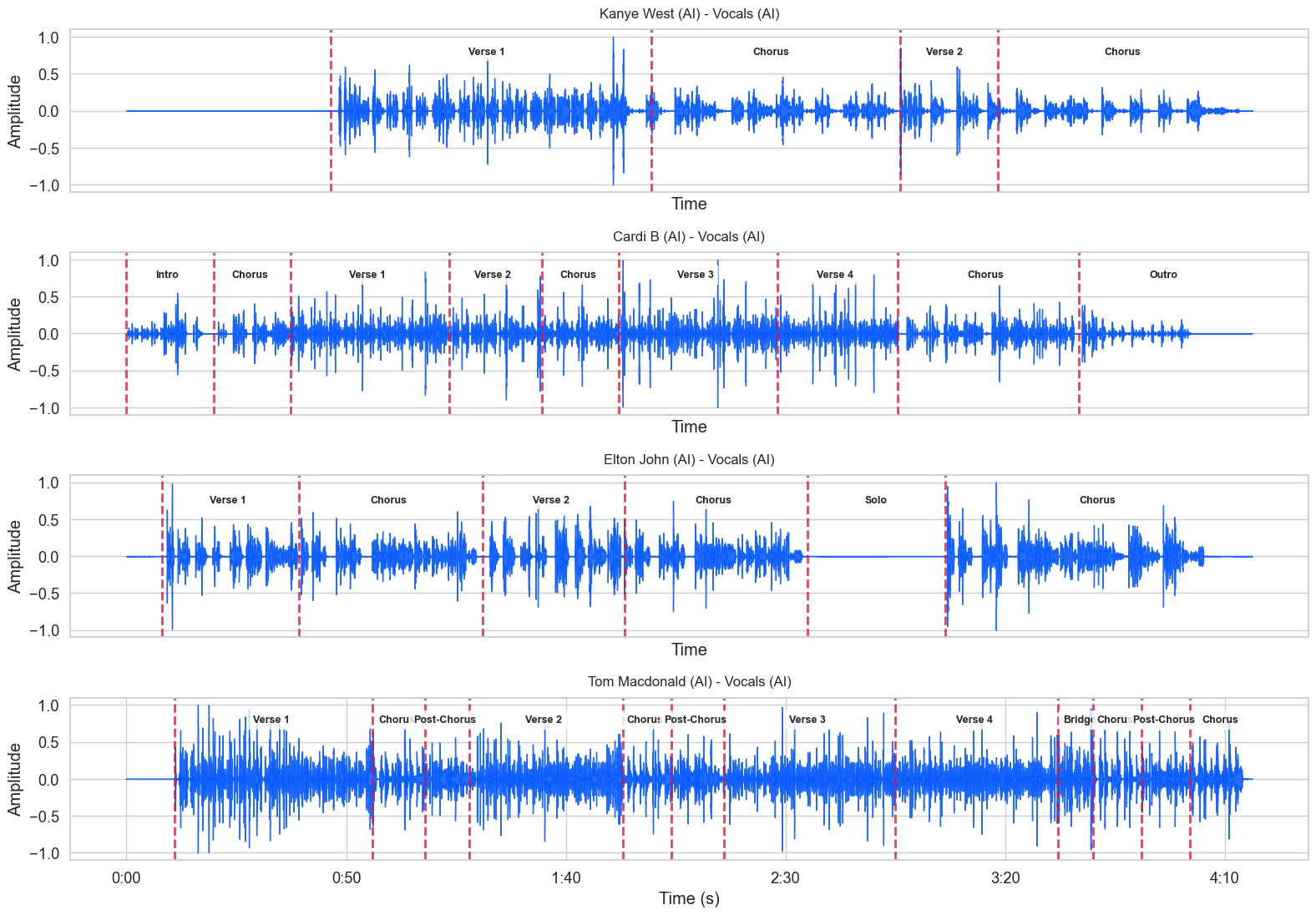}
    \caption{Waveform analysis by section after transformation}
    \label{fig:waveform_AI}
\end{figure*}

\begin{figure*}[htbp]
    \centering
    \includegraphics[width=1\linewidth]{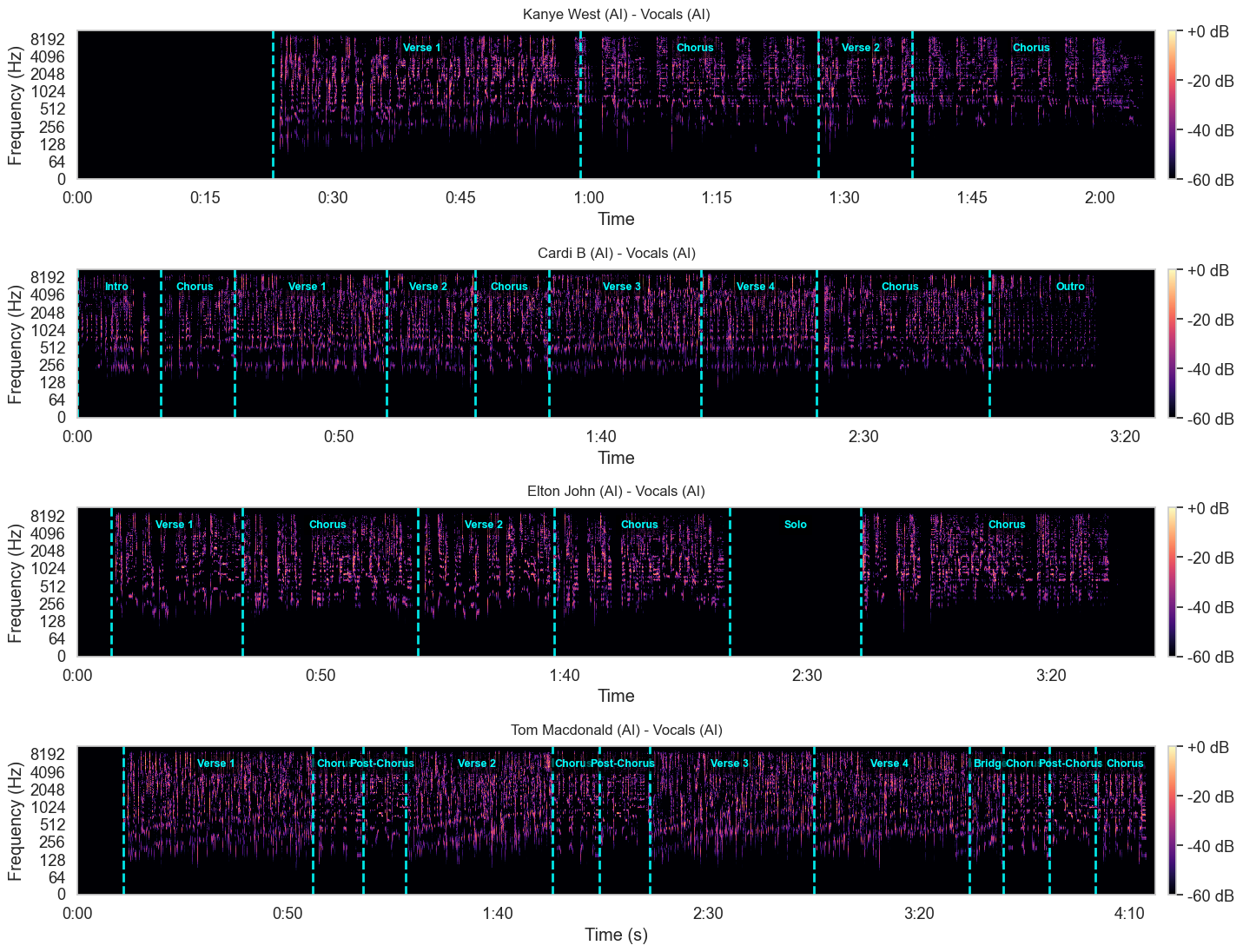}
    \caption{Spectrogram analysis divided by section after transformations}
    \label{fig:spectrogram_AI}
\end{figure*}

We compare the waveforms (Figure \ref{fig:waveform_AI}) of the original and transformed songs reveals subtle yet consistent differences. The transformed versions exhibit a more uniform loudness, while the originals display clearer transitions between sections and greater variation in intensity. The spectrogram analysis of Kanye West’s song (Figure \ref{fig:spectrogram_AI}) shows a reduction in overall intensity, indicated by the diminished presence of brighter regions in the higher frequency range. The transitions appear sharper but less impactful, with energy distributed more evenly across frequencies. Similarly, Cardi B’s transformation demonstrates reduced vocal sharpness and energy. Although the transformed spectrogram covers a similar frequency range as the original, it shows noticeably lower intensity, suggesting a smoother and less forceful vocal tone.

In contrast, Elton John’s transformed vocals display a marked increase in overall loudness across all song sections. The spectrogram reveals a more balanced frequency distribution, spanning from approximately 128 Hz to 8192 Hz. Tom MacDonald’s transformed track shows a similar waveform pattern to the original, with modest increases in amplitude. The corresponding spectrogram indicates a more even energy spread across a wider frequency range, reaching as low as 128 Hz. These results indicate that the Suno AI vocals exhibit a more balanced frequency and intensity compared to the original counterparts. Although some transformed tracks show mixed amplitude changes across sections, these variations alone do not strongly indicate aggressiveness.

We now utilise pitch-based analysis on the vocal stems for each artist (original vs transformed). Previous work has consistently demonstrated that certain features serve as reliable indicators of emotional states in human vocals. Juslin and Laukka  \cite{juslin2003}  identified that vocal anger and aggression are characterised by pitch variability, vocal intensity and spectral characteristics. Similarly, Bachorowski and Owren \cite{bachorowski} validated that aggressive vocal expressions show greater variability and vocal instability as measured by jitter and shimmer and Murray and Arnott \cite{arnott} related breathiness of vocals to anger and aggression.

We use features including HNR \cite{hnr}, CPP \cite{cpp}, Shimmer representing pitch instability, and Jitter reflecting amplitude instability, that define the key characteristics of the song. These four features allow for a nuanced examination of whether Suno AI-generated vocals can exhibit a less aggressive vocal quality than their human counterparts.

\begin{figure*}[htbp]
    \centering
    \includegraphics[width=0.75\linewidth]{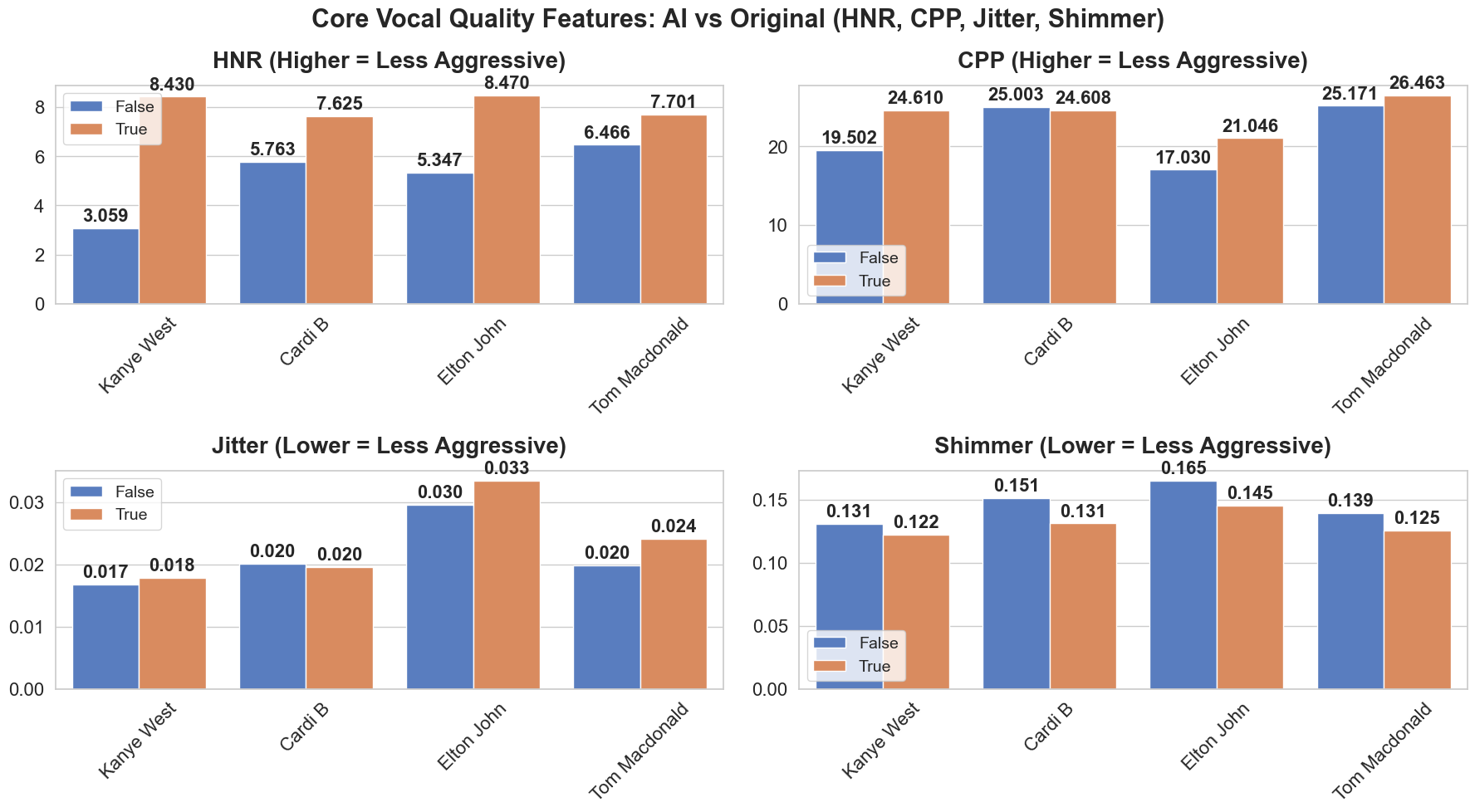}
    \caption{Bar plots before and after song transformation for HNR, CPP, Jitter, Shimmer.}
    \label{fig:audio}
\end{figure*}

\begin{figure*}[htbp]
    \centering
    \includegraphics[width=0.75\linewidth]{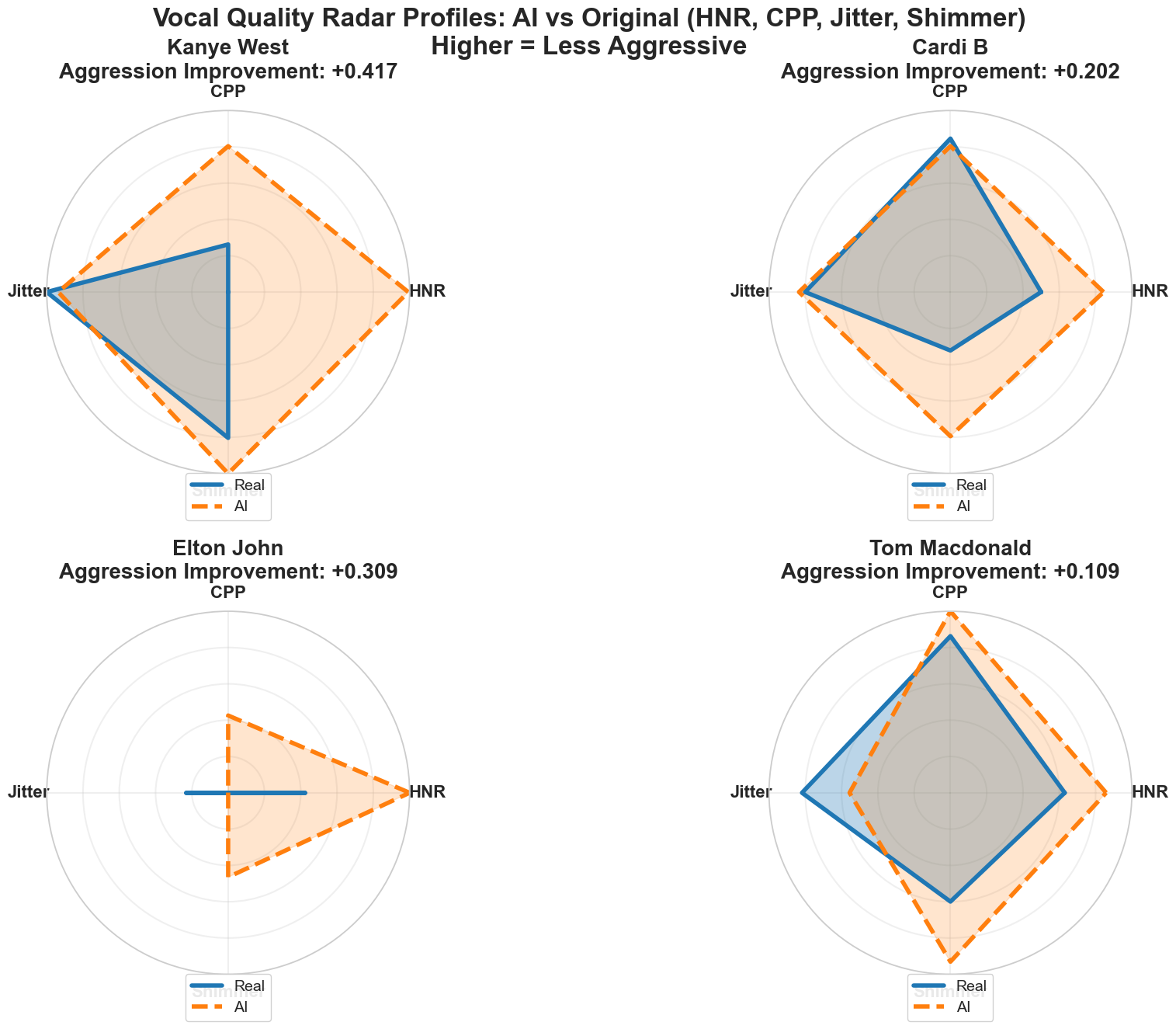}
    \caption{Original versus transformed song radar plot normalised between 0 and 1.}
    \label{fig:radar}
\end{figure*}

\begin{table}[htbp]
\centering
\caption{\textcolor{red}{RMS} Loudness and Range by Artist (Average volume of audio signal)}
\label{tab:rms_stats}
\begin{tabular}{l S[table-format=1.6] S[table-format=1.6] S[table-format=1.6]}
\toprule
\textbf{Artist} & \textbf{Avg RMS} & \textbf{Max RMS} & \textbf{Min RMS} \\
\midrule
\textbf{Human Artists} & & & \\
\midrule
Kanye West     & 0.213035 & 0.524201 & 0.000000 \\
Cardi B        & 0.227371 & 0.756325 & 0.000000 \\
Elton John     & 0.166824 & 0.345571 & 0.000080 \\
Tom Macdonald  & 0.263033 & 0.563769 & 0.000370 \\
\midrule
\textbf{AI Versions} & & & \\
\midrule
Kanye West (AI)     & 0.212540 & 0.456594 & 0.000001 \\
Cardi B (AI)        & 0.222747 & 0.429723 & 0.000001 \\
Elton John (AI)     & 0.226977 & 0.442575 & 0.000001 \\
Tom Macdonald (AI)  & 0.238674 & 0.460733 & 0.000001 \\
\bottomrule
\end{tabular}
\end{table}

We then analysed acoustic features from four artists (Kanye West, Cardi B, Elton John and Tom Macdonald) and compared them with Suno AI versions (Figure \ref{fig:audio}).

The HNR show higher values that indicate clearer and less noisy vocals, improved across all Suno-AI versions. Kanye West’s HNR increased from 3.06 dB to 8.43 dB, Cardi B’s from 5.76 dB to 7.62 dB, Elton John’s from 5.35 dB to 8.47 dB, and Tom Macdonald’s from 6.47 dB to 7.70 dB. These consistent increases suggest that Suno AI processing effectively reduced noise and enhanced harmonic clarity across all artists.

The CPP, a measure of vocal clarity and harmonic strength, also generally improved with Suno AI's vocals. Kanye West’s CPP increased from 19.50 to 24.61, Elton John’s from 17.03 to 21.05, and Tom Macdonald’s from 25.17 to 26.46, while Cardi B’s remained stable (from 25.00 to 24.61). Overall, this indicates that AI vocals tend to sound clearer and more resonant, though the effect size varies by artist.

The Jitter remained stable or slightly increased. Kanye West’s jitter rose marginally from 0.0168 to 0.0178, Cardi B’s decreased slightly from 0.0201 to 0.0195, Elton John’s increased from 0.0296 to 0.0334, and Tom Macdonald’s increased from 0.0198 to 0.0242. These small variations suggest that AI processing did not drastically alter pitch stability. The Shimmer was generally reduced, indicating smoother vocal dynamics. Kanye West’s shimmer decreased from 0.131 to 0.122, Cardi B’s from 0.151 to 0.131, Elton John’s from 0.165 to 0.145, and Tom Macdonald’s from 0.139 to 0.125. The consistent decrease across artists points to a cleaner, less harsh vocal output in the AI versions.

Based on the combined analysis of all four acoustic features (Figure \ref{fig:radar}), the findings indicate that Suno AI's vocal processing consistently produces vocals that are less aggressive and more acoustically stable. The transformation shows a clear trend toward vocal refinement (higher HNR), enhancing harmonic clarity (higher CPP), and improving both pitch and amplitude stability (lower Jitter and Shimmer) across most artists. Although the degree of improvement varies by performer, the overall pattern suggests that Suno AI's processing effectively smooths and clarifies vocal characteristics, resulting in cleaner and less harsh vocal outputs. We keep original and transformed MP3 files for the different songs with lyrics in our GitHub repository \footnote{\url{https://github.com/jchoi0406/musictransformationanalysis}}.

\section{Discussion}

Our research demonstrates that AI-driven vocal replacement represents a viable approach for reducing aggressiveness in music content. The framework successfully transformed four songs with explicitly aggressive content ("Heil Hitler," "WAP," "Island Girl," and "Whiteboy") into less aggressive versions while preserving core musical elements. We observed a consistent decrease in aggression in both lyrical sentiment analysis and audio feature examinations. 

The sentiment analysis revealed a consistent decrease in aggression scores ranging from 63.3\% to 85.6\% reduction, which confirmed the effectiveness of lyrical detoxification. Cardi B's 'WAP' showed the greatest improvement with each section reaching down to almost an 0\% aggression score. Similarly, Elton John's 'Island Girl' achieved a 73.2\% reduction. These results demonstrate AI's capacity to mitigate explicit content. However, we did see low semantic scores between original and transformed lyrics, especially with Tom Macdonald and Kanye West. This may mean LLMs struggle with certain types of content more than others, e.g. extreme politics vs racism and may also struggle when most lyrics are explicit, as seen with Kanye West's song.

The acoustic analysis complemented these findings, showing consistent vocal quality improvements. AI-generated vocals exhibited reduced pitch extremes, decreased vocal instability (shimmer), and improved harmonics-to-noise ratio (HNR). However, we did not see an improvement in Jitter. This parallel improvement in both lyrical content and vocal delivery creates a comprehensive mitigation of harmful content.

However, there are several limitations to our study. Our analysis is constrained to a small sample size (n=4), limiting generalisability across musical genres and aggression types. Technical limitations in current AI vocal replacement tools, particularly Suno AI's inconsistent beat preservation, introduced variability that may affect vocal-instrumental alignment. The non-deterministic nature of generative Suno AI's models also presents challenges for reproducible transformations.
Despite these limitations, our findings suggest that AI vocal replacement could enable safer listening experiences for both adults and children while preserving artistic intent more effectively than complete content removal. This approach aligns with evolving content moderation paradigms that prioritise transformation over censorship.

Several important directions emerge from this research. First, expanding the dataset to include more songs across diverse genres would strengthen the generalisability of our findings. Second, incorporating human perceptual studies would validate whether the acoustic and sentiment changes correspond to actual perceived aggression reduction. Third, improvements in AI vocal replacement, particularly better beat preservation and more consistent transformations, would enhance the framework's reliability. Finally, exploring real-time transformation systems could enable personalised content safety settings for streaming platforms.

\section{Conclusion}

This study demonstrates that AI-driven vocal replacement effectively reduces musical aggression through both lyrical detoxification and acoustic improvement. Our multi-modal analysis reveals consistent decreases in sentiment scores (63.3-85.6\%) alongside measurable vocal quality enhancements across all transformed songs. By reframing content moderation as a generative transformation task rather than simple filtering, we offer a nuanced approach that preserves artistic integrity while mitigating harmful content. The success of our framework across diverse artists suggests promising applications for safer listening experiences and responsible content moderation in music streaming platforms.

\section*{Code and Data}

Github repo \url{ https://github.com/jchoi0406/musictransformationanalysis}

 \bibliographystyle{elsarticle-num} 
 \bibliography{cas-refs}

\section*{Appendix}

\begin{table*}[htbp]
\centering
\caption{Chorus of original and transformed songs by artist}
\label{tab:chorus}
\begin{tabularx}{\textwidth}{l X X}
\toprule
\textbf{Artist} & \textbf{Original (Chorus + Timestamps)} & \textbf{AI (Chorus + Timestamps)} \\
\midrule

Kanye West &
\begin{tabular}[t]{@{}l@{}}
\textit{Timestamps: 0:51-1:16, 1:31-2:41} \\[0.5em]  
N*gga, Heil Hitler \\
N*gga, Heil Hitler (Hey) \\
They don't understand the things I say on Twitter \\
N*gga, Heil Hitler \\
They don't understand the things I say on Twitter \\
All my N*ggas godly, N*gga, Heil Hitler (Hey) \\
N*gga, Heil Hitler, N*gga, Heil Hitler \\
All my N*ggas Nazi's, N*gga, Heil Hitler
\end{tabular}
&
\begin{tabular}[t]{@{}l@{}}
\textit{Timestamps: 0:58-1:25, 1:38-2:16} \\[0.5em]  
Yeah, I rise higher \\
Yeah, I rise higher (Hey) \\
They don’t understand the pain I write on fire \\
Yeah, I rise higher \\
They don’t understand the pain I write on fire \\
All my people shining, yeah, we rise higher (Hey) \\
Yeah, we rise higher, yeah, we rise higher \\
All my people timeless, yeah, we rise higher
\end{tabular} \\
\hline

Cardi B &
\begin{tabular}[t]{@{}l@{}}
\textit{Timestamps: 0:15-0:28, 1:12-1:25, 2:18-2:45} \\[0.5em]  
Yeah, yeah, yeah, yeah \\
Yeah, you fuckin' with some wet-*ss p*ssy \\
Bring a bucket and a mop for this wet-*ss p*ssy \\
Give me everything you got for this wet-*ss p*ssy
\end{tabular}
&
\begin{tabular}[t]{@{}l@{}}
\textit{Timestamps: 0:29-0:42, 1:40-1:55, 2:47-3:14} \\[0.5em]  
Yeah, yeah, yeah, yeah \\
Yeah, you dealin' with some powerful women \\
Bring respect and your love for these powerful women \\
Give me everything you got for these powerful women
\end{tabular} \\
\hline

Elton John &
\begin{tabular}[t]{@{}l@{}}
\textit{Timestamps: 0:34-1:09, 1:43-3:42, 2:52-3:42} \\[0.5em]  
Island girl \\
What you wantin' with your white man's world? \\
Island girl \\
Black boy want you in his island world \\
He want to take you from your racket boss \\
He want to save you, but the cause is lost \\
Island girl, island girl, island girl \\
Tell me what you wantin' with your white man's world
\end{tabular}
&
\begin{tabular}[t]{@{}l@{}}
\textit{Timestamps: 0:34-1:09, 1:37-2:13, 2:42-3:41} \\[0.5em]  
Island girl \\
What you bringin’ to this bright new world? \\
Island girl \\
Proud boy want you in his island world \\
He want to lift you, watch you rise above \\
He want to guide you with a heart of love \\
Island girl, island girl, island girl \\
Tell me what you bringin’ to this bright new world
\end{tabular} \\
\hline

Tom Macdonald &
\begin{tabular}[t]{@{}l@{}}
\textit{Timestamps: 0:51-1:01, 1:48-1:58,}\\
\textit{3:52-4:31, 4:14-4:24} \\[0.5em]

White boy, don't say that \\
White boy, oh, you so bad \\
White boy, you wish you were black \\
White boy, dear white boy
\end{tabular}
&
\begin{tabular}[t]{@{}l@{}}
\textit{Timestamps: 0:56-1:07, 1:54-2:03,}\\
\textit{3:43-3:55, 4:07-4:30} \\[0.5em]  
Stand up, don't say that \\
Stand up, learn the facts \\
Stand up, use your voice back \\
Stand up, dear ally
\end{tabular} \\
\bottomrule
\end{tabularx}
\end{table*}

\end{document}